\documentclass[english,twoside,12pt,fleqn]{article}   
\evensidemargin=\oddsidemargin
\usepackage[parfill]{parskip}    	
\usepackage{graphicx}
\usepackage[utf8]{inputenc} 
\usepackage[english]{babel} 
\usepackage{extraipa}
\usepackage{amsmath,amsthm,amssymb}
\usepackage{array,hhline}
\usepackage{cite}    
\usepackage{color}
\usepackage[colorlinks,bookmarksdepth=2,linkcolor={blue},,urlcolor={red}]{hyperref}
\usepackage{caption}

\renewcommand{\emph}{\bfseries}
\newcommand{\ang}[1]{\raisebox{-6pt}{$\left#1\rule{0pt}{12pt}\right.$}}
\newcommand{\angShort}[1]{\raisebox{-2pt}{$\left#1\rule{0pt}{10pt}\right.$}}
 \newcommand{\angShortUp}[1]{\raisebox{-3pt}{$\left#1\rule{0pt}{11pt}\right.$}}
\newcommand{\angLong}[1]{\raisebox{-9pt}{$\left#1\rule{0pt}{16pt}\right.$}}
\newcommand{\angMed}[1]{\raisebox{-0.4pt}{$\left#1\rule{0pt}{22.98138pt}\right.$}}

\renewcommand{\Re}{\mathop{\mathrm{Re}}\nolimits}
\renewcommand{\Im}{\mathop{\mathrm{Im}}\nolimits}
\DeclareMathOperator{\arcosh}{arcosh}
\DeclareMathOperator{\arsinh}{arsinh}

\DeclareMathOperator{\rot}{\mathbf{rot}}

\allowdisplaybreaks[1]
\def\abstract{\topsep=0pt\partopsep=0pt\parsep=0pt\itemsep=0pt\relax 
\trivlist\item[\hskip\labelsep
{\bfseries\Large\abstractname}]\if!\abstractname!\hskip-\labelsep\fi\item[\hspace{0pt}]}

\providecommand{\keywords}[1]{\noindent \textit{Keywords:} #1}

 \title{\bf Scattering of Electromagnetic Waves Incident Normally on Square Patch-Type FSS Placed at the Interface between Two Dielectric Media}
 \date{}
\author{\\ \LARGE A.O. Tuzov \thanks{Department of Control systems, Siberian State Aerospace University, Krasnoyarsk,   Russia,  \hbox{e-mail: tuzov@sibsau.ru}}}

\begin{document}
\abovedisplayskip=6pt \belowdisplayskip=6pt

\maketitle
\begin{abstract}
In this paper an analytical approach for calculating scattering matrix elements for the case of normal incidence of the plane electromagnetic waves on the square patch-type Frequency Selective Surface (FSS), which is placed at the interface between two dielectric media is proposed. Analytical expressions for the reflection and transmission coefficients are shown to be accurate enough for practical purposes.


\end{abstract}

\section{Introduction}

At present, analytical and numerical methods~\cite{Mittra_71,Marcuvitz_86, Belaev_15,Belaev_17,Kontorovich_87, Tretyakov_03, Simovskii_08, Melchakova_Simovskii_08, Luukkonen_Simovskii_09, Ade_06,  Munk_00, Gupta_CAD_81, Gupta_microstrip_96, Mittra_13, Davidson_05} are widely used to solve problems of electromagnetic wave scattering by frequency selective surfaces (FSS).

In this paper an analytical approach for calculating scattering matrix elements for the case of normal incidence of plane electromagnetic waves on a square patch-type FSS from both sides, which is placed at an interface between two dielectric media with different dielectric permittivities, is proposed. Simple analytical expressions for the elements of the scattering matrix are derived under quasi-static assumption.

A comparison of frequency dependencies of the reflection and transmission coefficients calculated analytically by the derived formulae and computed numerically by 3D electromagnetic simulation is carried out.~\hspace{-1pt}Approximation error estimates of these analytical expressions~are~given.

\section{Analytical solution}
\subsection{Electromagnetic field}
\begin{figure}[h]	
	\includegraphics[scale=1]{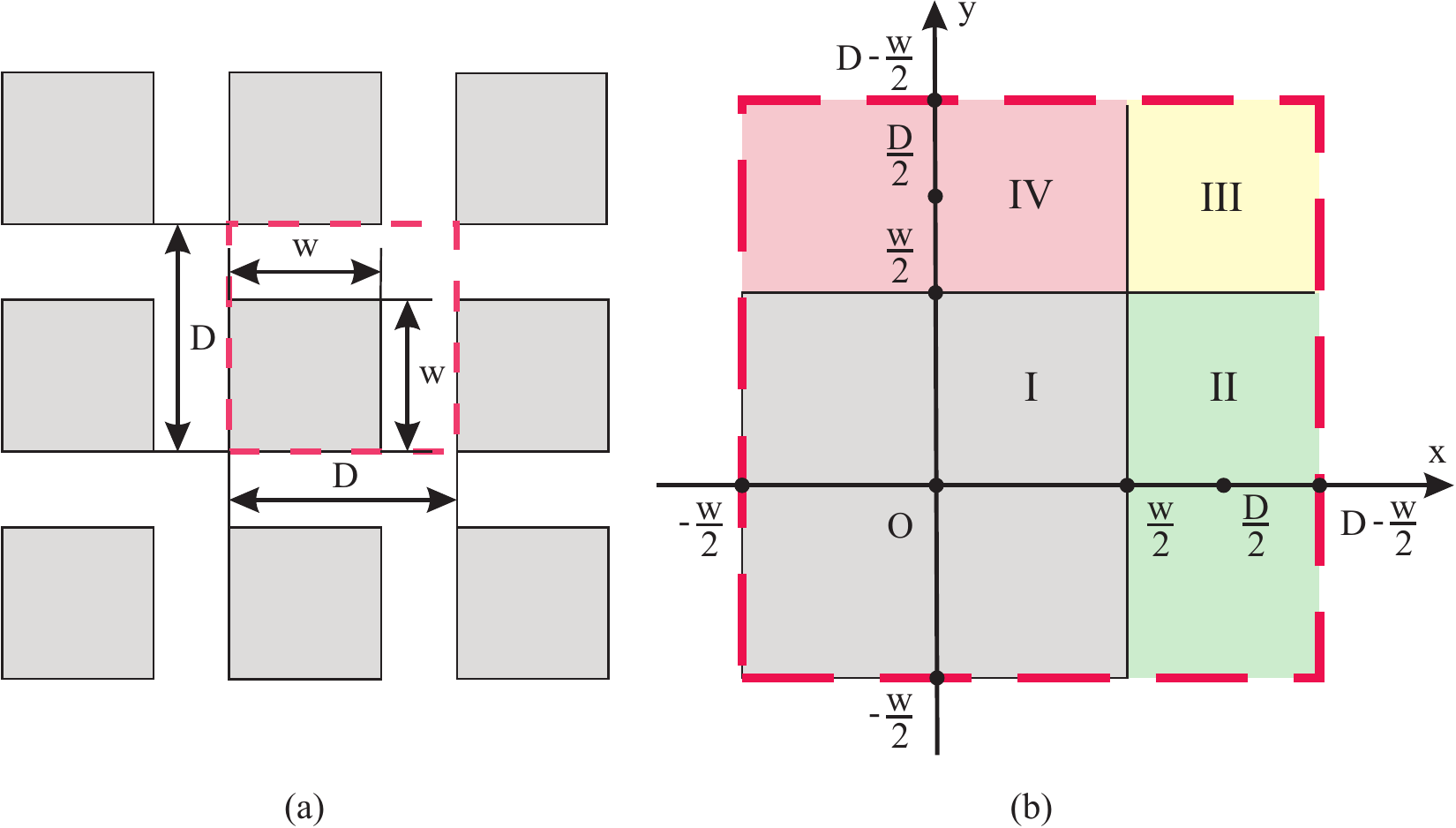}
	\caption{Two-dimensional view of patch-type FSS (a) and FSS unit cell (b). \\
	Region I for $z=0$ is an infinitesimally thin, perfectly conducting patch, \\
	regions II, III, IV -- a~dielectric medium homogeneous in the plane of the FSS ($z=0$) and  in~each~of~the~half-spaces:  
	relative dielectric permittivity equals $\varepsilon_{1}$ for $z<0$, $\varepsilon_{2}$ for $z \ge 0$.}
	\label{FIG1}
\end{figure}

Let us consider electromagnetic oscillations near the patch-type FSS (see Figure~ \ref{FIG1}) excited by two plane waves incident normally from both sides:
\begin{subequations}\label{E_H_Inc}
\begin{align}
E_x^{\mathrm{inc}}&=\left\{\begin{array}{cll}\label{E_Inc}
E_1^{\mathrm{inc}}& \exp \bigl({i{k_1}z - i\omega t}\bigr),	 \;\;\;\;\;\;\;  	&  z<0,\\
E_2^{\mathrm{inc}}& \exp \bigl({-i{k_2}z - i\omega t}\bigr),		& z>0,\\
\end{array}\right.\\
H_y^{\mathrm{inc}}&=\left\{\begin{array}{rll}\label{H_Inc}
\dfrac{E_1^{\mathrm{inc}}}{Z_1}		&	\exp \bigl({i{k_1}z - i\omega t}\bigr),		&  z<0, \vspace{6pt}\\
-\dfrac{E_2^{\mathrm{inc}}}{Z_2} 	&	\exp \bigl({-i{k_2}z - i\omega t}\bigr),		& z>0,\\
\end{array}\right.
\end{align}
\end{subequations}
where 
	$E_x^{\mathrm{inc}}, H_y^{\mathrm{inc}}$ are  component of electric and magnetic fields of the {\emph incident} waves, respectively;
	${k_{1,\, 2}} = \omega \sqrt {\mathstrut \varepsilon_{0} \,  \varepsilon_{1,2} \,  \mu_{0}}$ are wave numbers,
	${Z_{1,\, 2}} = Z_0/\sqrt{\mathstrut \varepsilon_{1,\, 2}}$ are characteristic impedance,
	$\varepsilon _{1,\, 2}$ are relative dielectric permittivities of medium 1 ($z<0$) and medium 2 ($z>0$), respectively separated by the FSS plane ($z=0$);
	$Z_0 = \sqrt {\mathstrut \mu_{0}/\varepsilon_{0}}$;
	$\varepsilon _0, \mu _0$ are absolute dielectric permittivity and absolute magnetic permeability of free space.

The field components of incident waves~\eqref{E_H_Inc} are homogeneous in the plane of the FSS ($E_x^{\mathrm{inc}}, H_y^{\mathrm{inc}}$ 
do not depend on $x, y$), hence according to Floquet's principle~\cite{Floquet_83} the components $E_x, H_y, E_z, H_z$ of the excited near-field of the FSS 
are periodic functions with~respect to $x, y$  with the period equal to the FSS unit cell size $D$. 
Therefore restrict ourself to consideration of the unit cell (Figure~ \ref{FIG1}~(b)).
Here $D$ is the unit cell size, $\mathrm w$ is the square patch width, $D-\mathrm w$ is the gap between patches. 

Under the quasi-static condition ($D\ll \lambda$, where $\lambda$ is a wavelength) the Helmholtz's equation is approximated by the Laplace's equation:
\begin{align}\label{eqs_Laplace}
\triangle E_x=0, \;\; \triangle H_y=0, \;\; \triangle E_z=0,  \;\;  \triangle H_z=0,
\end{align}
where $\triangle=\dfrac{\partial^{2}}{\partial x^{2}}+\dfrac{\partial^{2}}{\partial y^{2}}+\dfrac{\partial^{2}}{\partial z^{2}}$.

We find a solution of~\eqref{eqs_Laplace} as a linear combination of two particular solutions.
Since any function can be written as a sum of even and odd functions,  
we will find the component $E_x$ for the~first particular solution as an even function with~respect to $z$,
for the~second one -- as an odd function. Other components $H_y, E_z, H_z$ of these particular solutions are either even or odd functions with~respect to $z$.

\subsubsection{The first particular solution}
For the first particular solution of~\eqref{eqs_Laplace} we will find the component $E_x$ as an even function with~respect to $z$.
It follows from the time-harmonic Maxwell's equations that $E_x \propto (\rot{\mathbf H})_x$, $E_z \propto (\rot{\mathbf H})_z$.
Hence, $H_y$,  $E_z$ is odd, $H_z$ is even with~respect to $z$.

Asymptotic behavior of $E_x, E_z$ near edges is defined by the following edge conditions:
\begin{subequations}\label{edge_cond_E}
\begin{alignat}{2}
E_x \ang{|} \vphantom{E}_{
\begin{subarray}{1}
x=\mathrm w/2+\rho,\\ 
z = 0.
\end{subarray}} \propto
&E_x \angLong{|} \vphantom{E}_{
\begin{subarray}{1}
x=\mathrm w/2+\rho \cos \varphi ,\\
z=\rho \sin \varphi,\\
z \ne 0. 
\end{subarray}} \propto
\rho^{-1/2},&\qquad 
\rho \to +0.		\label{edge_cond_E_a}	\\
E_x \ang{|} \vphantom{E}_{
\begin{subarray}{1}
y=\mathrm w/2+\rho,\\ 
z = 0.
\end{subarray}} \propto
&E_x \angLong{|} \vphantom{E}_{
\begin{subarray}{1}
y=\mathrm w/2+\rho \cos \varphi ,\\
z=\rho \sin \varphi,\\
z \ne 0. 
\end{subarray}} \propto
\rho^{1/2},&\qquad 
\rho \to +0.\label{edge_cond_E_b}
\end{alignat}
\end{subequations}
where $\rho>0$ is the distance between the observation point $(x, y, z)$ and the nearest edge, $\rho$~is sufficiently small; distances to another edges are assumed to be much larger than $\rho$; $\varphi$ is the~polar angle, $-\pi < \varphi < \pi, \; \varphi \ne 0$;
the sign $\propto$ means asymptotic proportionality at $\rho \to +0$ (\mbox{$\Theta$-notation}).

The edge  conditions~\eqref{edge_cond_E} are the strengthening of Meixner's ones~\cite{Meixner_72, Mittra_71}.
Note that {\emph exact} solution for one-dimensional planar array FSS formed by infinitesimally thin parallel perfectly conducting strips~\cite{Belaev_15} satisfies~\eqref{edge_cond_E}.

It follows from~\eqref{edge_cond_E}  that in the regions I ($z \ne 0$) and II ($\forall z$) (see above~\ref{FIG1}(b)) near the edges ($\rho$ is sufficiently small): 
$\left | \dfrac{\partial^{2}E_x}{\partial x^{2}}\right | \gg \left | \dfrac{\partial^{2}E_x}{\partial y^{2}} \right | $.
Hence, in the Laplace's equation~\eqref{eqs_Laplace} the second derivative with~respect to $y$ can be neglected.
Therefore, we will obtain an analytical expression for  $E_x$,
in the regions I, II under the assumption of independence on~$y$.

At a sufficiently small distance from the edge $x= \mathrm w/2$ (so that distances to the another edges  $y= \pm\, \mathrm w/2$ are much larger than distance to $x= \mathrm w/2$)
the regions I and II can be approximately regarded as {\emph infinite along the y-axis} (I - an infinitesimally thin perfectly conducting strip,  \mbox{II - a~dielectric} medium).
In this case the incident waves~\eqref{E_H_Inc} are {\emph E-polarized} 
and electro-quasi-static approximation of Maxwell equations can be used, i.e.
 ${\mathbf E} = -\nabla  \widetilde{\varphi}$, 
where $ \widetilde{\varphi}$ is the electric scalar potential.
Due to the symmetry of ${\mathbf E}$ with~respect to the planes $x=0$, $z=0$ it is sufficient to restrict the computational domain to 
$\{(x,z):\quad 0 \le x \le D/2,\;  z \ge 0 \}$.
The~corresponding Laplace's equation for the electric potential is:
\begin{subequations}\label{BVP_phi}
\begin{align}
\triangle  \widetilde{\varphi}=0, \qquad  \triangle=\dfrac{\partial^{2}}{\partial x^{2}}+\dfrac{\partial^{2}}{\partial z^{2}},
\end{align}
with the boundary conditions:
\begin{align}
\widetilde{\varphi}&= \widetilde{\varphi}_1 \quad \text{ for } 0 < x \le  \mathrm w/2, \; z=0 \text{ and } x=0, \; z \ge 0,\\
\widetilde{\varphi}&= \widetilde{\varphi}_2 \quad \text{ for } x=D/2, \;  z \ge 0,
\end{align}
\end{subequations}
where $\widetilde{\varphi}_1,  \; \widetilde{\varphi}_2$ are some given constants.

The~exact solution of~\eqref{BVP_phi} obtained by analogy with~\cite{Belaev_15} by using the conformal mapping method, up to a constant multiplier, is: 
\begin{align*}
 \widetilde{\varphi}(x, z)=\Im V_0(z+i x),
\end{align*}
where $V_0(z+i x)=\arcosh\left(\dfrac{\cosh \left(\pi  (z+i x)/D\right)}{\cos \alpha}\right),  \quad \alpha= \dfrac{\pi \mathrm w}{2 D}$.

Hence,
\begin{equation}\label{E_x_E_z_I_II_t}
\begin{alignedat}{2}
E_x^{\mathrm{I,\, II}}(x, z) &=-\dfrac{\partial}{\partial x}\Im V_0(z+i x)	& & =\dfrac{\partial}{\partial x}\Im V(x+i z),\qquad z \ge 0, \\
E_z^{\mathrm{I,\, II}}(x, z) &=-\dfrac{\partial}{\partial z}\Im V_0(z+i x)	& & =\dfrac{\partial}{\partial z}\Im V(x+i z),\qquad z \ge 0,
\end{alignedat}
\end{equation}
since $V_0(z+i x)=\overline{V(x+i z)}$,
where  
\begin{alignat}{2}\label{V}
V(x+i z)=\arcosh\left(\dfrac{\cos \left(\pi  (x+i z)/D\right)}{\cos \alpha}\right), \qquad \alpha= \dfrac{\pi \mathrm w}{2 D},
\end{alignat}

Since in the first particular solution $E_x$ must be even function with~respect to $z$, we construct an even extension of $E_x$ to the half-space $z<0$:
\begin{align}\label{E_x_I_II}
E_x^{\mathrm{I,\, II}}(x, z)=\mp\, \dfrac{\partial}{\partial x}\Im V(x+i z),\qquad \forall z ,
\end{align}
where the upper sign ("-") corresponds to the half-space $z \le 0$,  the lower sign ("+") corresponds to the half-space $z \ge 0$; 
on the plane $z=0$ both expressions coincide.
Here  $\mp$ is due to the~oddness of the function $\dfrac{\partial}{\partial x}\Im V(x+i z)$ with~respect to~$z$.

Thus,  $E_x^{\mathrm{I,\, II}}$ satisfies the edge condition~\eqref{edge_cond_E_a}
and approximately satisfies (due to neglecting the~second derivative with~respect to $y$) the Laplace's equation in three dimensions~\eqref{eqs_Laplace}.

In analogy with~\eqref{E_x_I_II}, we derive from~\eqref{E_x_E_z_I_II_t} that:
\begin{align}\label{E_z_I_II}
E_z^{\mathrm{I,\, II}}(x, z)=\mp\, \dfrac{\partial}{\partial z}\Im V(x+i z),\qquad \forall z.
\end{align}
The $z$-component of the time-harmonic Maxwell's equations:
$(\rot{\mathbf H})_z=-i \,  \omega \,  \varepsilon_{0} \, \varepsilon _{1,\, 2}E_z,  \; \forall z$, therefore
\begin{align*}
\dfrac{\partial H_y^{\mathrm{I,\, II}}}{\partial x}=\pm\, i \,  \omega \,  \varepsilon_{0} \, \varepsilon _{1,\, 2} \dfrac{\partial}{\partial z}\Im V(x+i z).
\end{align*}
The function $V(x+i z)$ defined by~\eqref{V}  is an~analytic function in the regions under consideration ($-\mathrm w/2<x<D-\mathrm w/2, \, z<0$ and $-\mathrm w/2<x<D-\mathrm w/2, z>0$)
(in each half-spaces $z<0$, $z>0$ separately).
Hence, its real and imaginary parts, considered as functions of two real variables, satisfy the~Cauchy–Riemann equations in these regions, so that
\begin{align}\label{H_y_dx}
\dfrac{\partial H_y^{\mathrm{I,\, II}}}{\partial x}=\pm\, i \,  \omega \,  \varepsilon_{0} \, \varepsilon _{1,\, 2} \dfrac{\partial}{\partial x}\Re V(x+i z).
\end{align}
Integrating~\eqref{H_y_dx} with~respect to $x$ yields
\begin{align}\label{H_y_I_II}
H_y^{\mathrm{I,\, II}}(x, z)=\pm\, i \,  \omega \,  \varepsilon_{0} \,   \dfrac{\varepsilon_{1}+\varepsilon_{2}}{2}  \Re V(x+i z).
\end{align}
Here an additive {\emph function of integration} is defined so that $H_y$ is odd function with~respect to $z$, as the first particular solution requires.

It follows from boundary conditions for the electromagnetic field at the planar interface between two dielectric media that 
$H_y$ in the regions II, III, IV is continuous across the interface~($z=0$): 
\begin{align*}
H_y \angShort{|} \vphantom{H}_{
\begin{subarray}{1}
\vphantom{z}\\
z =-0
\end{subarray}
} =
H_y \angShort{|} \vphantom{H}_{
\begin{subarray}{1}
\vphantom{z}\\
z =+0
\end{subarray}
} .
\end{align*}
Continuity and oddness of $H_y$ with~respect to $z$ results in
\begin{align}\label{H_z_0_cond}
H_y^{\mathrm{II, \, III, \, IV}} \angShort{|} \vphantom{H}_{
\begin{subarray}{1}
\vphantom{z}\\
z =\mp\, 0
\end{subarray}
}=0. 
\end{align}
Note that $H_y^{\mathrm{ II}}$ defined by~\eqref{H_y_I_II} satisfies the interface condition~\eqref{H_z_0_cond}.

It follows from~\eqref{H_z_0_cond},~\eqref{H_y_I_II},~\eqref{V} that
\begin{align}\label{H_y_I_II_avg} 
\hspace{-\mathindent} 
\langle H_y  \rangle\stackrel{\mathrm{def}}{=}  \dfrac{1}{D^2} \int\limits_{-\mathrm w/2}^{D-\mathrm w/2} \,  \int\limits_{-\mathrm w/2}^{D-\mathrm w/2} H_y \angShort{|} \vphantom{H}_{\begin{subarray}{1} \vphantom{z}\\z = \mp\, 0\end{subarray}} \, dx \, dy
=\dfrac{\mathrm w}{D^2} \int\limits_{-\mathrm w/2}^{\mathrm w/2} H_y^{\mathrm{I}} \angShortUp{|} \vphantom{H}_{\begin{subarray}{1} \vphantom{z}\\z = \mp\, 0\end{subarray}} \, dx
=\mp\, i \,  \omega \,  \varepsilon_{0} \,   \dfrac{\varepsilon_{1}+\varepsilon_{2}}{2}  {\mathrm w}/D \ln \cos \alpha, \quad \hspace{-2em} 
\end{align}
since 
\begin{align*}
\hspace{-\mathindent} & \int\limits_{-\mathrm w/2}^{\mathrm w/2} \Re V(x+i z) \angShortUp{|} \vphantom{H}_{\begin{subarray}{1} \vphantom{z}\\z = \mp\, 0\end{subarray}} \, dx=
	 \Re  \int\limits_{-\mathrm w/2}^{\mathrm w/2} \arcosh\left(\dfrac{\cos \left(\pi  x/D\right)}{\cos \alpha}\right) \, dx=
	 \dfrac{D}{\pi} \Re  \int\limits_{-\alpha}^{\alpha} \arcosh\left(\dfrac{\cos t}{\cos \alpha}\right) dt=\\
\hspace{-\mathindent} & =-D  \ln \cos \alpha. 
\end{align*}

Let us obtain an analytical expression for  $E_x$ in the regions I ($\forall z$), IV ($\forall z$). 
For this purpose, we first derive analytical expressions for $H_y$.

Asymptotic behavior of $H_y$ near the edge $y=\mathrm w/2$ is defined by:
\begin{subequations}\label{edge_cond_H}
\begin{align} \label{edge_cond_H_a} 
H_y \ang{|} \vphantom{H}_{
\begin{subarray}{1}
y=\mathrm w/2-\rho,\\ 
z = 0.
\end{subarray}} \propto
H_y \angLong{|} \vphantom{H}_{
\begin{subarray}{1}
y=\mathrm w/2+\rho \cos \varphi ,\\
z=\rho \sin \varphi,	\\
z \ne 0. 
\end{subarray}} \propto
\rho^{-1/2},\qquad 
\rho \to +0,
\end{align}
and near  the edge $x=\mathrm w/2$ by:
\begin{align}\label{edge_cond_H_b}
H_y \ang{|} \vphantom{H}_{
\begin{subarray}{1}
x=\mathrm w/2-\rho,\\ 
z = 0.
\end{subarray}} \propto
H_y \angLong{|} \vphantom{H}_{
\begin{subarray}{1}
x=\mathrm w/2+\rho \cos \varphi ,\\
z=\rho \sin \varphi,\\
z \ne 0. 
\end{subarray}} \propto
\rho^{1/2},\qquad 
\rho \to +0.		
\end{align}
\end{subequations}
The edge  conditions~\eqref{edge_cond_H} are the strengthening of Meixner's ones~\cite{ Meixner_72, Mittra_71}.
Note that {\emph exact} solution for one-dimensional planar array FSS formed by infinitesimally thin parallel perfectly conducting strips~\cite{Belaev_15} satisfies~\eqref{edge_cond_H}.

It follows from~\eqref{edge_cond_H}  that in the Laplace's equation~\eqref{eqs_Laplace} in the regions I ($\forall z$) and IV ($z \ne 0$) near the edges ($\rho$ is sufficiently small): 
$\left | \dfrac{\partial^{2}H_y}{\partial y^{2}}\right | \gg \left | \dfrac{\partial^{2}H_y}{\partial x^{2}} \right | $.

Hence,  dependence of $H_y$ on $x$ is weak (compared with dependence on $y$)  in the regions I,  IV ($\forall z$) (here, at $z=0$, consistently with~\eqref{H_z_0_cond}, $H_y^{\mathrm{IV}}=0$). 

At a sufficiently small distance from the edge $y= \mathrm w/2$ (so that distances to the another edges  $x= \pm\, \mathrm w/2$ are much larger than distance to $y= \mathrm w/2$)
the regions I and IV can be approximately regarded as {\emph infinite along the x-axis} (I - an infinitesimally thin perfectly conducting strip, II - a~dielectric medium).

In this case the incident waves~\eqref{E_H_Inc} are {\emph H-polarized} 
and magneto-quasi-static approximation of Maxwell equations can be used, i.e.
${\mathbf H} = -\nabla \psi$, 
where $\psi$ is the magnetic scalar potential.
Due to the symmetry of ${\mathbf H}$ with~respect to the planes $y=0$, $z=0$ it is sufficient to restrict the computational domain to 
$\{(y, z):\quad 0 \le y \le D/2,\;  z \ge 0 \}$.
The~corresponding Laplace's equation for the magnetic potential is:
\begin{subequations}\label{BVP_psi}
\begin{align}
\triangle  \psi=0, \qquad  \triangle=\dfrac{\partial^{2}}{\partial y^{2}}+\dfrac{\partial^{2}}{\partial z^{2}},
\end{align}
with the boundary conditions:
\begin{align}
\psi&= \psi_1 \quad \text{ for }  y=0, \; z \ge 0,\\
\psi&= \psi_2 \quad \text{ for } \mathrm w/2 \le y <  D/2, \; z=0 \text{ and } y=D/2, \;  z \ge 0,
\end{align}
\end{subequations}
where $\psi_1,  \; \psi_2$ are some given constants.

The~exact solution of~\eqref{BVP_psi} obtained by analogy with~\cite{Belaev_15} by using the conformal mapping method, up to a constant multiplier, is: 
\begin{align*}
 \psi(y, z)=\Im \Psi_0(z+i y),
\end{align*}
where 
$\Psi_0(z+i y)=\arsinh\left(\dfrac{\sinh \left(\pi  (z+i y)/D\right)}{\sin \alpha}\right),  \quad \alpha= \dfrac{\pi \mathrm w}{2 D}$.

Hence,
\begin{alignat}{2}\label{H_y_I_IV_t}
H_y^{\mathrm{I,\, IV}}(y, z) &=-\dfrac{\partial}{\partial y}\Im \Psi_0(z+i y) & =-\dfrac{\partial}{\partial y}\Re \Psi(y+i z),\qquad z \ge 0,
\end{alignat}
since $\Psi_0(z+i y)=i \, \overline{\Psi(y+i z)}$,
where  
\begin{alignat}{2}\label{Psi}
\Psi(y+i z)=\arcsin\left(\dfrac{\sin \left(\pi  (y+i z)/D\right)}{\sin \alpha}\right), \qquad \alpha= \dfrac{\pi \mathrm w}{2 D},
\end{alignat}

Thus,  $H_y^{\mathrm{I,\, IV}}(y, z)$ defined by~\eqref{H_y_I_IV_t} satisfies only the edge condition~\eqref{edge_cond_H_a}
and {\emph exactly} satisfies  the corresponding {\emph two}-dimensional Laplace's equation (\eqref{eqs_Laplace} 
without the~second derivative with~respect to~$x$).

As we have remarked, dependence of $H_y$ on $x$ is weak in the regions I ($\forall z$),  IV ($\forall z$), hence 
$H_y$ can be constructed by combining~\eqref{H_y_I_IV_t} and~\eqref{H_y_I_II} in a manner similar to~\cite{Belaev_17}.
\begin{align*}
H_y(x, y, z)= C  \Re V(x)  \dfrac{\partial}{\partial y}\Re \Psi (y+i z), \qquad z \ge 0,
\end{align*}
where $C$ is an undetermined constant multiplier.

Since in the first particular solution $H_y$ must be odd function with~respect to $z$, we construct an odd extension of $H_y$ to the half-space $z<0$:
\begin{align}\label{H_y_C}
H_y(x, y, z)=\pm\, C  \Re V(x)  \dfrac{\partial}{\partial y}\Re \Psi (y+i z), \qquad \forall z.
\end{align}
Here $\pm$ is due to the evenness of the function $\dfrac{\partial}{\partial y}\Re \Psi(y+i z)$ with~respect to~$z$.
Furthermore, $H_y(x, y, z)$ satisfies the condition~\eqref{H_z_0_cond}.

Thus, $H_y(x, y, z)$ defined by~\eqref{H_y_C} in the regions I ($\forall z$), IV ($\forall z$)
is valid also in II ($z=0$), III ($z=0$) consistently with~\eqref{H_z_0_cond}.
Here, $H_y(x, y, z)$, in contrast to $H_y^{\mathrm{I,\, IV}}(y, z)$, satisfies both~\eqref{edge_cond_H_a} and \eqref{edge_cond_H_b} edge conditions and
approximately satisfies (due to the smallness of the second derivative with~respect to $x$) the Laplace's equation in {\emph three} dimensions~\eqref{eqs_Laplace}.

To determine the constant $C$, calculate $\langle H_y  \rangle$ again:
\begin{align}\label{H_y_avg}
\langle H_y  \rangle = \pm\, C  \dfrac{1}{D^2}    \Re \Psi (y \mp\, i\, 0) \angMed{|}_{-\mathrm w/2}^{\mathrm w/2}  \; \cdot \;   \int\limits_{-\mathrm w/2}^{\mathrm w/2}\Re V(x) \, dx 
= \mp \, C \dfrac{\pi}{D} \ln \cos \alpha,
\end{align}
since
\begin{align*} 
\hspace{-\mathindent} 
 \int\limits_{-\mathrm w/2}^{\mathrm w/2}\Re V(x) \, dx = -D  \ln \cos \alpha, \quad
\Re \Psi (y \mp\, i\, 0) \angMed{|}_{-\mathrm w/2}^{\mathrm w/2}  \hspace{-16pt}  =\Re \bigl[\arcsin(1 \mp\, i \, 0) - \arcsin(-1 \mp\, i \, 0)\bigr ] = \pi.
\end{align*} 
Comparing~\eqref{H_y_avg} with~\eqref{H_y_I_II_avg}, we obtain
\begin{align*}
C= i \,  \omega \,  \varepsilon_{0} \,   \dfrac{\varepsilon_{1}+\varepsilon_{2}}{2}  \dfrac{\mathrm w}{\pi},
\end{align*}
then~\eqref{H_y_C} takes the form:
\begin{align}\label{H_y}
H_y(x, y,z)=\pm\, i \,  \omega \,  \varepsilon_{0} \,   \dfrac{\varepsilon_{1}+\varepsilon_{2}}{2}  \dfrac{\mathrm w}{\pi}  \Re V(x)  \dfrac{\partial}{\partial y}\Re \Psi (y+i z).
\end{align}

\enlargethispage{\baselineskip}
The $y$-component of the time-harmonic Maxwell's equations:
$(\rot{\mathbf E})_y=i \,  \omega  \mu_{0} H_y, \; \forall z$, \hbox{therefore}
\begin{align*}
\dfrac{\partial E_x}{\partial z}=
\mp\, \omega^{2} \,  \varepsilon_{0} \,  \dfrac{\varepsilon_{1}+\varepsilon_{2}}{2} \mu_{0} \dfrac{\mathrm w}{\pi}  \Re V(x)  \dfrac{\partial}{\partial y}\Re \Psi (y+i z).
\end{align*}
The function $\Psi (y+i z)$ defined by~\eqref{Psi}  is an~analytic function in the regions under consideration.
Hence, its real and imaginary parts, considered as functions of two real variables, satisfy the~Cauchy–Riemann equations, so that
\begin{align}\label{E_x_dz}
\dfrac{\partial E_x}{\partial z}=
\mp\, \omega^{2} \,  \varepsilon_{0} \,  \dfrac{\varepsilon_{1}+\varepsilon_{2}}{2} \mu_{0} \dfrac{\mathrm w}{\pi}  \Re V(x)  \dfrac{\partial}{\partial z}\Im \Psi (y+i z).
\end{align}
Integrating~\eqref{E_x_dz} with~respect to $z$ yields in the regions I ($\forall z$), IV ($\forall z$)
\begin{align}\label{E_x}
E_x(x, y,z)=\mp\, \omega^{2} \,  \varepsilon_{0} \,  \dfrac{\varepsilon_{1}+\varepsilon_{2}}{2} \mu_{0} \dfrac{\mathrm w}{\pi}  \Re V(x) \Im \Psi (y+i z).
\end{align}
Here an additive function of integration is defined from interface conditions for the electromagnetic field.

It follows from~\eqref{E_x_I_II},~\eqref{E_x}  that 
\vspace{6pt}
\begin{equation}\label{E_x_avg}
\begin{alignedat}{3}
\langle E_x  \rangle &\stackrel{\mathrm{def}}{=}\dfrac{1}{D^2} \int\limits_{-\mathrm w/2}^{D-\mathrm w/2} \,  \int\limits_{-\mathrm w/2}^{D-\mathrm w/2} E_x \angShort{|} \vphantom{H}_{\begin{subarray}{1} \vphantom{z}\\z = \mp\, 0\end{subarray}} \, dx \, dy=
\dfrac{\mathrm w}{D^2} \int\limits_{\mathrm w/2}^{D-\mathrm w/2} E_x^{\mathrm{II}}(x, z)  \angShortUp{|} \vphantom{E}_{\begin{subarray}{1} \vphantom{z}\\z = \mp\, 0\end{subarray}} \, dx+\\
&+\dfrac{1}{D^2} \int\limits_{\mathrm w/2}^{D-\mathrm w/2} \,  \int\limits_{-\mathrm w/2}^{\mathrm w/2} E_x(x, y,z)  \angShortUp{|} \vphantom{H}_{\begin{subarray}{1} \vphantom{z}\\z = \mp\, 0\end{subarray}} \, dx \, dy
= -\dfrac{\pi \mathrm w}{D^2}  +\omega^{2} \,  \varepsilon_{0} \,   \dfrac{\varepsilon_{1}+\varepsilon_{2}}{2} \mu_{0}  \dfrac{\mathrm w}{\pi} \ln \cos \alpha  \cdot  \ln \sin \alpha,	
\end{alignedat} \hspace{-4em} 
\end{equation}

since
\vspace{12pt}
\begin{align*}
\int\limits_{\mathrm w/2}^{D-\mathrm w/2} E_x^{\mathrm{II}}(x, z)  \angShortUp{|} \vphantom{E}_{\begin{subarray}{1} \vphantom{z}\\z = \mp\, 0\end{subarray}} \, dx &=
\mp\,  \bigl{[} \Im V(D-\mathrm w/2 \mp\, i \, 0)-\Im V(\mathrm w/2 \mp\, i \, 0) \bigr{]}=\\
&=\mp\,  \Im \bigl{[}  \arcosh(-1 \pm\, i \, 0)-  \arcosh(1 \pm\, i \, 0) \bigr{]}=- \pi,
\end{align*}
\begin{align*}
\int\limits_{-\mathrm w/2}^{\mathrm w/2} \Re V(x) \angShortUp{|} \vphantom{H}_{\begin{subarray}{1} \vphantom{z}\\z = \mp\, 0\end{subarray}} \, dx &=
\Re  \int\limits_{-\mathrm w/2}^{\mathrm w/2} \arcosh\left(\dfrac{\cos \left(\pi  x/D\right)}{\cos \alpha}\right) \, dx=
\dfrac{D}{\pi} \Re  \int\limits_{-\alpha}^{\alpha} \arcosh\left(\dfrac{\cos t}{\cos \alpha}\right) dt=\\
& =-D  \ln \cos \alpha, 
\end{align*}
\begin{align*}
&\int\limits_{\mathrm w/2}^{D-\mathrm w/2} \Im \Psi (y+i z)   \angShortUp{|} \vphantom{H}_{\begin{subarray}{1} \vphantom{z}\\z = \mp\, 0\end{subarray}} \, dy =
\mp\, \int\limits_{\mathrm w/2}^{D-\mathrm w/2} \arcosh\left(\dfrac{\sin \left(\pi  y/D\right)}{\sin \alpha}\right) \, dy =\\
&=\mp\, \dfrac{D}{\pi} \int\limits_{\alpha}^{\pi -\alpha} \arcosh\left(\dfrac{\sin t}{\sin \alpha}\right) \, dt =
\mp\, \dfrac{D}{\pi} \int\limits_{-\beta}^{\beta} \arcosh\left(\dfrac{\cos s}{\cos \beta}\right) \, ds =
\pm\, D \ln \cos \beta = \pm\, D \ln \sin \alpha,
\end{align*}
where $s=t-\pi/2$, $\beta=\pi/2-\alpha$.

\subsubsection{The second particular solution}
Components of the second particular solution of~\eqref{eqs_Laplace} are supplementary to ones of the first particular solution, 
i.e. $E_x$, $H_z$ are odd functions,  $H_y$,  $E_z$ are even functions with~respect to~$z$. 
The simplest nontrivial solution of~\eqref{eqs_Laplace} satisfying these requirements is
\begin{align}\label{solution2}
E_x=0, \;\; H_y=H_{0}, \;\; E_z=0, \;\; H_z=0,
\end{align}
where $H_{0} \ne 0$ is some constant.

Thus, electric and magnetic field components $ \widetilde{E}_x,  \widetilde{H}_y$, 
which are a linear combination of the~first and second particular solutions with the coefficients \hfil $c_{1}$,\enskip  $c_{2}$,\hfil 
have the following average  \hbox to 0pt{values:}
\begin{align}\label{E_H_tilde_avg}
\langle \widetilde{E}_x  \rangle = c_{1} \langle E_x  \rangle, \quad \langle  \widetilde{H}_{y}  \rangle = c_{1} \langle H_y  \rangle+c_{2} H_{0}
\end{align}
respectively,\enskip where $\langle E_x  \rangle, \langle H_y  \rangle$ are defined by~\eqref{E_x_avg},~\eqref{H_y_I_II_avg}.

\subsection{Scattering matrix}
The considered  patch-type FSS can be modelled as a two-port network:
some domains to the left ($z<0$) and right  ($z>0$) of the FSS plane ($z=0$) is viewed as port 1 and 2 respectively.
The~relationship between the incident and scattered waves is described by scattering matrix~$\mathbf{S}$:
\begin{align}\label{sys_S_matrix}
\mathbf{b}=\mathbf{S\, a},
\end{align}
where $\mathbf{a}=(a_{1}, a_{2})^{\intercal}$, $\mathbf{b}=(b_{1}, b_{2})^{\intercal}$.

Here $a_{1}, a_{2}$	are the normalized complex amplitudes of the~incident waves and $b_{1}, b_{2}$ are ones of the scattered waves
at port 1 and 2 respectively~\cite{Gupta_CAD_81}:
\begin{subequations}\label{a_b_via_E_H_inc_sct}
\begin{align}\label{a_b_via_E_inc_sct}
a_{1,\, 2}=E_{1,\, 2}^{\mathrm{inc}} / \sqrt { Z_{1,\, 2}}\: , \quad b_{1,\, 2}=E_{1,\, 2}^{\mathrm{sct}} / \sqrt {Z_{1,\, 2}} \:,
\end{align}
where $E_{1,\, 2}^{\mathrm{inc}}$\enskip are the complex amplitudes of the electric fields of the incident waves~\eqref{E_Inc},
$E_{1,\, 2}^{\mathrm{sct}}$\enskip ~are the  complex amplitudes of the electric fields of the scattered waves at port 1 and 2 \hbox{respectively.}

It follows from~\eqref{a_b_via_E_inc_sct} that
\begin{align}\label{a_b_via_H_inc_sct}
H_{1,\, 2}^{\mathrm{inc}}=\pm\, a_{1,\, 2} / \sqrt { Z_{1,\, 2}}\:, \quad H_{1,\, 2}^{\mathrm{sct}}= \mp\, b_{1,\, 2} / \sqrt { Z_{1,\, 2}},
\end{align}
where the upper and lower signs correspond to the port 1 and 2 respectively;
$H_{1,\, 2}^{\mathrm{inc}}$, $H_{1,\, 2}^{\mathrm{sct}}$\enskip are the complex amplitudes of the magnetic fields of the incident~\eqref{H_Inc} and scattered waves \hbox{respectively.}
\end{subequations}

Boundary conditions for the electromagnetic field at ports 1 and 2 are of the form
\begin{equation}\label{E_H_boundary}
\begin{alignedat}{2}
E_{1,\, 2}^{\mathrm{inc}}\, +\, &E_{1,\, 2}^{\mathrm{sct}}	& &= \langle \widetilde{E}_x  \rangle,\\
H_{1,\, 2}^{\mathrm{inc}}\, +\, &H_{1,\, 2}^{\mathrm{sct}}	& &= \langle \widetilde{H}_y  \rangle,
\end{alignedat}
\end{equation}
where $ \langle \widetilde{E}_x  \rangle, \langle \widetilde{H}_y  \rangle$ are defined by~\eqref{E_H_tilde_avg}.

Substituting~\eqref{E_H_tilde_avg},~\eqref{a_b_via_E_H_inc_sct} into~\eqref{E_H_boundary} gives
\begin{equation}\label{sys_a_b}
\begin{alignedat}{3}
&a_{1,\, 2}\sqrt { Z_{1,\, 2}}				& + 	\,&b_{1,\, 2}\sqrt {Z_{1,\, 2}}			& &= c_{1} \langle E_x \rangle\\
\pm\, &a_{1,\, 2} / \sqrt { Z_{1,\, 2}} 	&\enskip \mp\, &b_{1,\, 2} / \sqrt { Z_{1,\, 2}}		& &= c_{1} \langle H_y  \rangle+c_{2} H_{0},
\end{alignedat}
\end{equation}
where $\langle E_x  \rangle, \langle H_y  \rangle$ are defined by~\eqref{E_x_avg},~\eqref{H_y_I_II_avg}.

Let us rewrite the system of linear algebraic equations~\eqref{sys_a_b} in the form~\eqref{sys_S_matrix}. 
Denote \linebreak $\widehat{Z}=\langle E_x  \rangle /  \langle H_y  \rangle$, 
where $\langle E_x  \rangle$,  $\langle H_y  \rangle$ are evaluated, for the sake of definiteness, at~port~1.
Then
\begin{equation}\label{sys_a_b_v2}
\begin{aligned}
b_1 =  \dfrac{\sqrt {\mathstrut \varepsilon_{1}}  - \sqrt {\mathstrut \varepsilon_{2}}  -  Z_0/Z}{\sqrt {\mathstrut \varepsilon_{1}} + \sqrt {\mathstrut \varepsilon_{2}}  +  Z_0/Z} \; a_1
+\dfrac{2 \sqrt[4]{\mathstrut \varepsilon_{1} \varepsilon_{2}}}{\sqrt {\mathstrut \varepsilon_{1}} + \sqrt {\mathstrut \varepsilon_{2}}  +  Z_0/Z} \: a_2, \\[6pt]
b_2 =  \dfrac{2 \sqrt[4]{\mathstrut \varepsilon_{1} \varepsilon_{2}}}{\sqrt {\mathstrut \varepsilon_{1}} + \sqrt {\mathstrut \varepsilon_{2}}  +  Z_0/Z} \; a_1
+\dfrac{\sqrt {\mathstrut \varepsilon_{2}}  - \sqrt {\mathstrut \varepsilon_{1}}  -  Z_0/Z}{\sqrt {\mathstrut \varepsilon_{1}} + \sqrt {\mathstrut \varepsilon_{2}}  +  Z_0/Z} \: a_2.\\[6pt]
\end{aligned}
\end{equation}

\vspace{6pt}
Substituting $Z= i\, X$ into~\eqref{sys_a_b_v2} and using~\eqref{E_x_avg},~\eqref{H_y_I_II_avg}, we obtain
\vspace{18pt}
\begin{subequations}\label{S_matrix}
\begin{align}\label{S}
\mathbf S  =\dfrac{1}{\sqrt {\mathstrut \varepsilon_{1}} + \sqrt {\mathstrut \varepsilon_{2}}  -i\, Z_0/X}
    \begin{pmatrix}
   \sqrt {\mathstrut \varepsilon_{1}}  - \sqrt {\mathstrut \varepsilon_{2}}  + i\, Z_0/X & 2 \sqrt[4]{\mathstrut \varepsilon_{1} \varepsilon_{2}}  \\ 
   2 \sqrt[4]{\mathstrut \varepsilon_{1} \varepsilon_{2}}& {\sqrt {\mathstrut \varepsilon_{2}} - \sqrt {\mathstrut \varepsilon_{1}} + i\, Z_0/X}  
    \end{pmatrix}, 
\end{align}
\vspace{6pt}
\begin{align}\label{X}
X =\omega \, \mu_{0} \,  \dfrac{D}{2 \pi}   \,   \ln \sin \alpha	-	\dfrac{\pi}{\omega \, \varepsilon_{0} (\varepsilon_{1} + \varepsilon_{2})D}  \,  \dfrac{1}{ \ln \cos \alpha}\, , 
\end{align}
\end{subequations}

\vspace{6pt}
where $\alpha = \dfrac{\pi \mathrm w}{2 D}, \;\; Z_0 = \sqrt {\mathstrut \mu_{0}/\varepsilon_{0}}\, .$

\vspace{18pt}
Electromagnetic wave propagation can be described using a transmission line equivalent circuit model.
Then $Z$ is  the normalized electrical impedance of the unit cell of the patch-type FSS,
X~is the normalized electrical reactance, 
the first term in~\eqref{X} is inductive reactance,
the~second one is  capacitive reactance.

\clearpage
\section{Numerical solution and comparison of the results}
Let us estimate the approximation error of the formulae derived in this paper. For this purpose, the absolute values of the reflection coefficient $S_{1 1}$ and the transmission coefficient $S_{2 1}$ calculated by the analytical expression~\eqref{S_matrix}, have been compared with ones computed numerically with high accuracy by 3D electromagnetic simulation with CST MWS. Note that $S_{2 1}=S_{1 2}$ and $|S_{1 1}|=|S_{2 2}|$.

\begin{figure}[h]
	\includegraphics[scale=1]{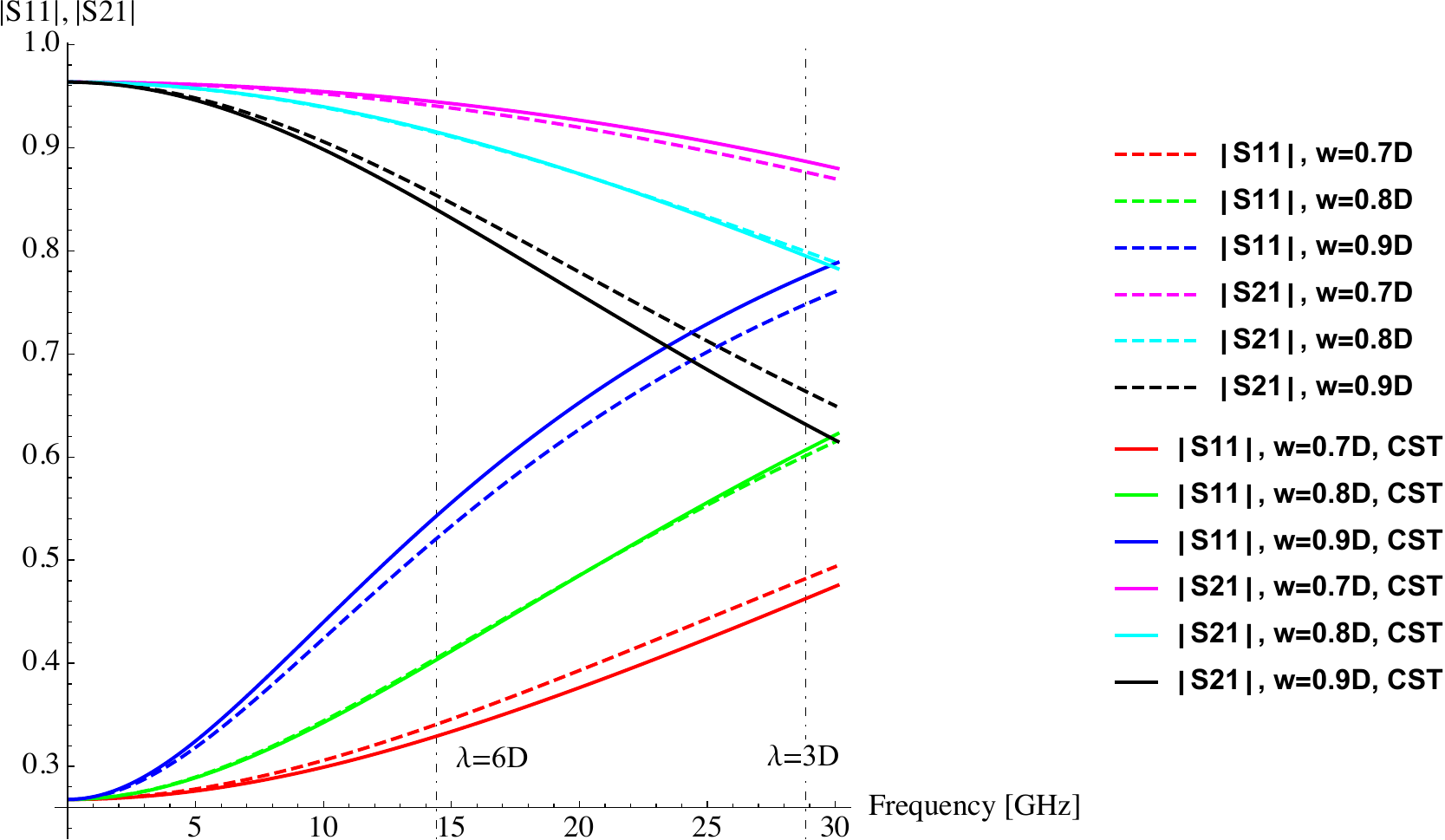}
	\caption{Frequency dependencies of the absolute values of the reflection coefficient $S_{1 1}$ and transmission coefficient $S_{2 1}$ 
	of  the electromagnetic waves incident normally on the patch-type FSS with the fixed period $D$=2 mm, 
	the variable relative width of the patch ${\mathrm w}/D=0.7, 0.8, 0.9$ and $\varepsilon_1=1$,  $\varepsilon_2=3$.
	\newline
Dashed lines correspond to the analytical solution calculated by the formula~\eqref{S_matrix}, solid lines represent a numerical solution obtained with high accuracy by 3D electromagnetic simulation with CST MWS.}
\label{FIG2}
\end{figure}

Figure~\ref{FIG2} shows frequency dependencies of $|S_{1 1}|, |S_{2 1}|$ of the electromagnetic waves incident normally on the patch-type FSS with the fixed period $D$
and the variable relative width of the patch ${\mathrm w}/D$.
\enlargethispage{\baselineskip}

The reflection coefficients increase and the transmission coefficients decrease with increasing the relative width of the patch, as displayed in Figure~\ref{FIG2}.

\begin{figure}[h]
	\includegraphics[scale=1]{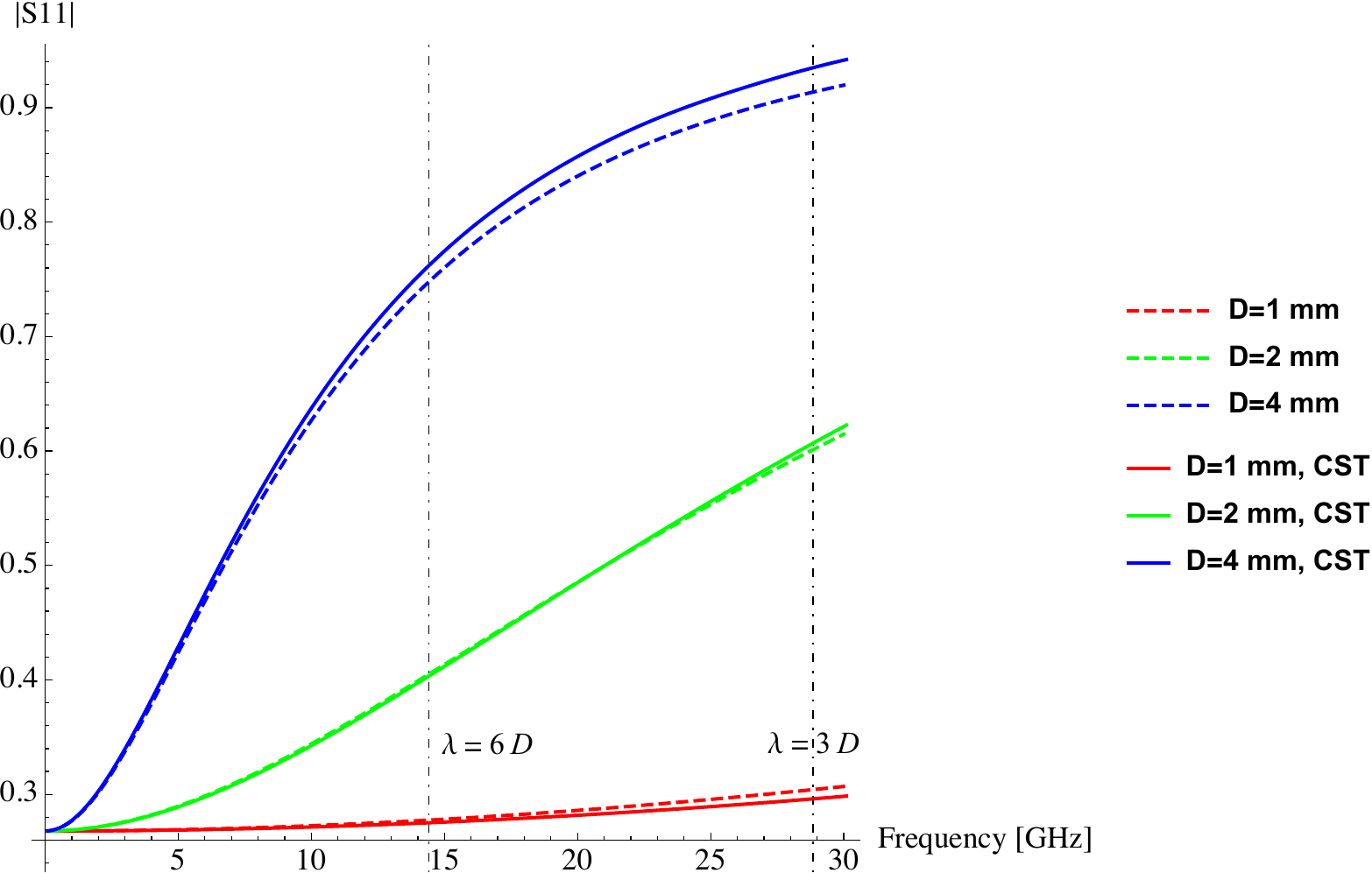}
	\caption{Frequency dependencies of the absolute value of the reflection coefficient $S_{1 1}$  of  the~electromagnetic waves incident normally on the patch-type FSS with the fixed gap between the patches \mbox{$D-{\mathrm w}=0.4$}, the variable period $D=1, 2, 4$ mm and $\varepsilon_1=1$,  $\varepsilon_2=3$.
	\newline	
	Dashed lines correspond to the analytical solution calculated by the formula~\eqref{S_matrix}, solid lines represent a numerical solution obtained with high accuracy by 3D electromagnetic simulation with CST MWS.}
	\label{FIG3}
\end{figure}

Figure~\ref{FIG3} shows frequency dependencies of $|S_{1 1}|$  of  the electromagnetic waves incident normally on the patch-type FSS 
with the fixed gap between the patches $D-{\mathrm w}$ and the variable period~$D$.

The reflection coefficients increase with increasing the patch-type FSS period for the fixed gap between the patches, as displayed in Figure~\ref{FIG3}.

Recall that the approximate analytical solution has been obtained under $\lambda \gg D$, therefore, as expected,  
the approximation error for $\lambda=6 D$ is smaller than for $\lambda=3 D$, as shown in~Figures~\ref{FIG2},~\ref{FIG3}. 
Here $\lambda$	is the wavelength in the second medium ($\varepsilon_2=3$).

Table~\ref{TAB1} presents the estimation of the relative approximation error of the analytical expressions for different frequencies.
\begin{table}[h]
	\caption{Frequency and wavelength dependencies of the relative deviations $\delta S_{11}$,  $\delta S_{21}$ of  the~reflection  and transmission coefficients $S_{11}$,  $S_{21}$ calculated by the analytical expression~\eqref{S_matrix} from $S_{11}^{\text{CST}}$, $S_{21}^{\text{CST}}$ obtained with high accuracy by 3D electromagnetic simulation with CST MWS for $D=2$ mm, ${\mathrm w}=0.8 D$, $\varepsilon_1=1$,  $\varepsilon_2=3$;
$\lambda$	is the wavelength in the second medium ($\varepsilon_2=3$). Here $\delta S_{ij}=|S_{ij}-S_{ij}^{\text{CST}}|/|S_{ij}^{\text{CST}}|$.}
\begin{tabular}{|c|c|c|c|}
\hline
\multicolumn{1}{| m{32 mm} |}{Frequency [GHz]}  & \multicolumn{1}{| m{32 mm} |}{Wavelength [mm] }  & \multicolumn{1}{c|}{$\delta S_{11}$}& \multicolumn{1}{c|}{$\delta S_{21}$}\\
\hline
2 & 86.5 & $4.1 \times 10^{-4}$ & $3.2 \times 10^{-5}$\\
\hline
4 & 43.3 & $1.4 \times 10^{-3}$ & $1.2 \times 10^{-4}$\\
\hline
6 & 28.8 & $2.7 \times 10^{-3}$ & $2.6 \times 10^{-4}$\\
\hline
8 & 21.6	& $3.8 \times 10^{-3}$ & $4.3 \times 10^{-4}$\\
\hline
10 & 17.3	& $4.5 \times 10^{-3}$ & $6.0 \times 10^{-4}$\\
\hline
12 & 14.4	& $4.7 \times 10^{-3}$ & $7.4 \times 10^{-4}$\\
\hline
14 & 12.4	& $4.3 \times 10^{-3}$ & $8.0 \times 10^{-4}$\\
\hline
16 & 10.8	& $3.4 \times 10^{-3}$ & $7.5 \times 10^{-4}$\\
\hline
\end{tabular}
\label{TAB1}
\end{table}

\vspace{-6pt}
The relative deviations $\delta S_{11}$ and $\delta S_{21}$ does not exceed $0.5\:  \%$ and $0.08\:  \%$ respectively
in the~frequency range up to $\widehat{f}=$ 16 GHz, i.e. in the wavelength range down to 10.8 mm ($\lambda > 6 D$), as~shown in Table~\ref{TAB1}.

\section{Conclusion}
Thus, in this paper the simple, but quite accurate analytical expressions for the elements of the scattering matrix have been derived under the quasi-static assumption for the case of normal incidence of the plane electromagnetic waves on the square patch-type FSS from both sides, which is placed at the interface between two dielectric media with the different dielectric permittivities.

The comparison of frequency dependencies of the reflection and transmission coefficients calculated analytically by the derived formulae and computed numerically with high accuracy by 3D electromagnetic simulation with CST MWS has shown good agreement between both approaches. Numerical results have demonstrated that the formulae obtained in this paper are accurate enough for practical purposes in their applicability domain. 

The derived analytical expressions can be used in design of multi-layer patch-type FSS structures. 	They can help to {\emph analytically} optimize the FSS structure parameters and hence avoid extensive numerical simulations, and therefore reduce computational costs. Desired reflective properties of such structures can be achieved by varying both the relative width of the patch ${\mathrm w}/D$ and the FSS period $D$.


\begin{thebibliography}{999}
\bibitem{Floquet_83}
M. G. Floquet, “Sur les equations diffkrentielles IinCaires a coefficients pCriodiques,” Annale E‘coleNormale Siiperieur, pp. 47-88, 1883.

  \bibitem{Meixner_72} 
J. Meixner, The behavior of electromagnetic fields at edges,  IEEE Transactions on Antennas and Propagation, Vol. 20, No. 4, 442-446, 1972.

  \bibitem{Mittra_71} 
R. Mittra, S. W. Lee, Analytical Techniques in the Theory of Guided Waves. New York, NY: MacMillan, 1971.
  
  \bibitem{Marcuvitz_86} 
N. Marcuvitz, Waveguide Handbook, Electromagnetic waves series //IEEE Peter Peregrinus, London. - 1986. -T. 21. - C. 298.

  \bibitem{Ade_06} 
P.A.R. Ade , G. Pisano , C. Tucker , S. Weaver, A Review of Metal Mesh Filters  // Proc. SPIE. 2006. V. 6275. P. 62750U-1.

  \bibitem{Belaev_15} 
B.A. Belyaev,  V.V. Tyurnev, Diffraction of electromagnetic waves on a one-dimensional strip conductor grating located at the interface between dielectric media, Russian Physics Journal, Vol. 58, No 5,  646-657, 2015.

  \bibitem{Belaev_17} 
B.A. Belyaev,  V.V. Tyurnev, Scattering of electromagnetic waves on a metal grating located at the interface between dielectric media, Journal of Radio Electronics (JRE), No. 7, 2017.

\bibitem{Munk_00}
B. Munk, Frequency selective surfaces: theory and design. N.Y.: John Wiley \& Sons, Inc., 2000.

\bibitem{Gupta_CAD_81}
K.C. Gupta, R. Garg, R. Chadha, Computer aided design of microwave circuits, Dedham, MA: Artech House, 1981.

\bibitem{Gupta_microstrip_96}
K.C. Gupta, R. Garg, I. J. Bahl, Microstrip Lines and Slotlines, second edition, Artech House Inc., Dedham, MA, 1996.

\bibitem{Kontorovich_87}
M.I. Kontorovich, M.I. Astrakhan, V.P. Akimov, G.A. Fersman, Electrodynamics of Grid Structures, Moskov: Radio i Svyaz, 1987.

\bibitem{Tretyakov_03}
S. Tretyakov, Analytical modeling in applied electromagnetics, Norwood, MA: Artech House, 2003.

\bibitem{Simovskii_08}
K.R. Simovskii A.A. Sochava, I.V. Melchakova, A high impedance surface with stable low resonance frequency, Journal of Communication Technology and Electronics, Vol. 53, 497-507, 2008.

\bibitem{Melchakova_Simovskii_08}
I.V. Melchakova, K.R. Simovskii, Efficient simple analytic model of artificial impedance surfaces based on resonance microstrip grids, Journal of Communication Technology and Electronics, Vol. 53, 874-884, 2008.

\bibitem{Luukkonen_Simovskii_09}
O. Luukkonen, P. Alitalo, C.R. Simovski, and S.A. Tretyakov, Experimental verification of an analytical model for high-impedance surfaces, Electronics Letters, vol. 45, no. 14, pp. 720-721, 2009.

\bibitem{Mittra_13}
R. Mittra  (ed.), Computer Techniques for Electromagnetics: International Series of Monographs in Electrical Engineering. – Elsevier, 2013.~-T.~7.

\bibitem{Davidson_05}
D. B. Davidson,  Computational electromagnetics for RF and microwave engineering. – Cambridge University Press, 2005.

\end{thebibliography}
\end{document}